\documentclass[preprint,aps,eqsecnum]{revtex4-1}

\usepackage{amsmath,amssymb,amsfonts,dcolumn,color,graphicx,graphics,
latexsym,placeins,epsfig}
\usepackage{subfigure,rotating,hyperref}

\newcommand{\be}{\begin{equation}}
\newcommand{\ee}{\end{equation}}
\newcommand{\ba}{\begin{eqnarray}}
\newcommand{\ea}{\end{eqnarray}}

\usepackage[normalem]{ulem}
\newcommand{\stkout}[1]{\ifmmode\text{\sout{\ensuremath{#1}}}\else\sout{#1}\fi}

\begin{document}

\title{\Large \bf Gravitational collapse in $(2+1)$-dimensional Eddington-inspired Born-Infeld gravity}

\author{Rajibul Shaikh ${}^{1,}$} 
\email{rajibul.shaikh@tifr.res.in}
\author{Pankaj S. Joshi ${}^{1,2,}$}
\email{psj@tifr.res.in} 
\affiliation{${}^{1}$ Tata Institute of Fundamental Research, Homi Bhabha Road, Colaba, Mumbai 400005, India}
\affiliation{${}^{2}$ International Center for Cosmology, Charusat University, Anand 388421, Gujarat, India}
\bigskip

\begin{abstract} 
We study here the gravitational collapse of dust in $(2+1)$-dimensional 
spacetimes for the formation of black holes (BH) and naked singularities (NS)
as final states in a modified theory of gravity, with vanishing cosmological constant. From the perspective of cosmic censorship, we investigate the collapse of a dust cloud in  Eddington-inspired Born-Infeld gravity (EiBI) and compare the results with those of general relativity (GR). It turns out that, as opposed to the general relativistic situation, where the outcome of dust collapse in $(2+1)$ dimensions is always a naked singularity, the EiBI theory has a certain range of parameter values that avoid the naked singularity. This indicates that a $(3+1)$-dimensional generalization of these results could be useful and worth examining. Finally, using the results here, we show that the singularity avoidance through homogeneous bounce in cosmology in this modified gravity is not stable.

\end{abstract}

\maketitle

\section{Introduction}
Several  studies  on  gravitational  collapse  scenarios  in (2+1) dimensional 
spacetimes  have  been  carried  out by  various  authors  \citep{LDC1,LDC2,LDC3,LDC4,LDC5}.  
These  provide  interesting toy  models  which  may give  important  insights
from the perspective of quantum gravity. This is because in (2+1) dimensional 
spacetimes there is no gravity outside the matter. Also the spacetime metric is always conformally flat as the Weyl tensor  vanishes  identically  everywhere.
On the other hand, the situation in (3+1) and higher dimensions is far more
complicated compared to this \citep{LDC5,Joshi}.

To investigate the final collapse outcome in (2+1) dimensional collapse, let 
us consider the geometry of trapped surfaces in this case. In general, for a 
spherically symmetric spacetime, the equation for apparent horizon is given 
by $g^{\mu\nu}\partial_\mu R\partial_\nu R=0$, where $g_{\mu\nu}$ is the metric tensor, and $R$ is the area radius. In GR, the equation of an apparent horizon can alternatively be expressed as $F/R^{N-3}=1$, where $N$ is the dimension of the spacetime, and $F$ is the mass function for the collapsing cloud \citep{LDC5}. 
Thus, we see that the equation of apparent horizon in the (2+1) case is given 
by $F(t,r)=1$ for the gravitational collapse of any general matter field. 

It is therefore interesting to note that, in (2+1)-dimensional GR, the geometry of trapped 
surfaces is completely determined by the mass function alone of the cloud, 
and is independent of the area radius of the collapsing shells.  If the  mass  function  
of  the  collapsing configuration is  bounded  from above, with say 
$F (t,r)<1$ for the range $-\infty < t < t_s$ and $0 < r < r_b$, where $t_s$
and $r_b$ are the singularity time and the boundary of the cloud respectively,
we then see that the trapping does not occur during the collapse and the 
complete  singularity that forms as the collapse end state is necessarily 
visible to an outside observer. This situation is is strikingly different from four or higher-dimensional 
spherically symmetric spacetimes, where  a  massive  singularity  is always 
necessarily trapped within an apparent horizon.

A very interesting subcase of this situation is that of $(2+1)$
dimensional  dust  collapse in GR,  where $F(t, r) = F (r)$ gives the equation for 
the apparent horizon, with  the mass  function having no time dependence
in this case \citep{LDC5}. Here we see that the  initial  mass  of  the  collapsing  cloud  
completely determines the final outcome in terms of BH or NS. Clouds
with small enough mass always form a visible singularity, whereas for larger 
masses a trapped region is present at all epochs.  However,  as  demanded  
by  the  regularity  conditions, we must avoid trapped surfaces on the initial 
surface $t = t_i$, where the collapse commences. In that case, for the
dust collapse case, there are no trapped surfaces developing  at  all  at  
any other  later  epochs  until  the  singularity formation, and as a result 
the $(2+1)$ dimensions dust collapse always necessarily  produces  
a  visible  naked  singularity (NS),  as  opposed to the BH/NS phases obtained 
in usual four-dimensional dust collapse.

However, the situation may be different in modified theories of gravity. In general, most of the modified gravity theories significantly differ from GR in strong curvature regimes. Therefore, in such regimes, the dynamics of gravitational collapse, the formation of spacetime singularities, trapped surfaces, the formation and dynamics of apparent horizons etc, may significantly differ from that found in GR. As a simplest scenario, we investigate the collapse of a dust cloud with zero cosmological constant in (2+1) dimension in a modified theory of gravity, namely the  Eddington inspired Born-Infeld gravity (EiBI) \citep{banados1} and compare the results with those of GR. The EiBI gravity theory is a class of Born-Infeld inspired gravity theory first suggested by Deser and Gibbons \citep{deser}, suggested by the earlier work of Eddington \citep{eddington}, and the nonlinear electrodynamics of Born and Infeld \citep{born_infeld}. The EiBI theory is equivalent to Einstein's GR in vacuum but differs from it in the presence of matter. Since its introduction, various aspects of EiBI gravity have been studied by many researchers in the recent past. The final fate of a gravitational collapse of homogeneous dust in EiBI gravity was studied in \cite{EiBIc}. Various aspects such as black holes \citep{banados1,EiBIBH}, wormholes \citep{EiBIWH}, compact stars \citep{EiBICS}, cosmological aspects \citep{banados1,delsate1,EiBICOS,cho1}, astrophysical aspects \citep{EiBIASTRO}, gravitational waves \citep{EiBIGW} etc. have been worked out. See \cite{EiBIreview} for a recent review on various studies in EiBI gravity. Our main purpose here is to study the effect of introducing inhomogeneities in matter, towards the collapse final states.

\section{Eddington-inspired Born-Infeld gravity}
\label{sec:EiBI}
The action in EiBI gravity is given by
\begin{eqnarray}
S_{BI}[g,\Gamma,\Psi]&=&\frac{2}{\kappa}\int d^3x\left[\sqrt{-\left\vert g_{\mu\nu}+\kappa R_{\mu\nu}(\Gamma)\right\vert}-\lambda \sqrt{-\left\vert g\right\vert}\right]+S_{M}(g,\Psi),
\end{eqnarray}
where $\lambda=1+\kappa\Lambda$, $\kappa$ is the EiBI theory parameter, $R_{\mu\nu}(\Gamma)$ is the symmetric part of the Ricci tensor built with the independent connection $\Gamma$,  $S_{M}(g,\Psi)$ is the action for the matter field, $\Lambda$ is the cosmological constant, and the vertical bars stand for the matrix determinant. Variations of this action with respect to the metric tensor $g_{\mu\nu}$ and  the connection $\Gamma$ yield, respectively, \citep{banados1,cho1,delsate1}
\begin{equation}
\sqrt{-q}q^{\mu\nu}=\lambda \sqrt{-g}g^{\mu\nu}-\kappa \sqrt{-g}T^{\mu\nu}
\label{eq:field_equation1}
\end{equation}
\begin{equation}
\nabla^\Gamma_\alpha \left(\sqrt{-q} q^{\mu\nu}  \right)=0,
\label{eq:metric_compatibility}
\end{equation}
where $\nabla^\Gamma$ denotes the covariant derivative defined by the 
connection $\Gamma$, and $q^{\mu\nu}$ is the inverse of the auxiliary metric 
$q_{\mu\nu}$ defined by
\begin{equation}
q_{\mu\nu}=g_{\mu\nu}+\kappa R_{\mu\nu}(\Gamma).
\label{eq:field_equation2}
\end{equation}
In obtaining the field equations from the variation of the action, it is assumed that both the connection $\Gamma$ and the Ricci tensor $R_{\mu\nu}(\Gamma)$ are symmetric, i.e., $\Gamma^\mu_{\nu\rho}=\Gamma^\mu_{\rho\nu}$ and $R_{\mu\nu}(\Gamma)=R_{\nu\mu}(\Gamma)$. Equation (\ref{eq:metric_compatibility}) gives the metric compatibility equation which yields
\begin{equation}
\Gamma^\mu_{\nu\rho}=\frac{1}{2}q^{\mu\sigma}\left(q_{\nu\sigma,\rho}+q_{\rho\sigma,\nu}-q_{\nu\rho,\sigma} \right).
\end{equation}
Therefore, the connection $\Gamma^\mu_{\nu\rho}$ is the Levi-Civita connection of the auxiliary metric $q_{\mu\nu}$. Either in vacuum or in the limit $\kappa\to 0$, GR is recovered \citep{banados1}.

In this work, we study inhomogeneous dust collapse in $(2+1)$ dimensions. It has been shown that, in $(3+1)$ dimensions, the field equations (\ref{eq:field_equation1}) and (\ref{eq:field_equation2}) can be combined to write an effective Einstein field equation $G^{\mu}_\nu[q]+\Lambda\delta^{\mu}_\nu=\frac{1}{\lambda}\mathcal{T}^{\mu}_\nu$ for the auxiliary metric $q_{\mu\nu}$ \citep{delsate1}, where $\mathcal{T}^{\mu}_\nu$ is an apparent energy-momentum tensor dependent on the physical energy-momentum tensor $T^\mu_{\nu}$. To show this for an arbitrary $N$ dimension, we first rewrite Eqs. (\ref{eq:field_equation1}) and (\ref{eq:field_equation2}), respectively, as
\begin{equation}
q^{\mu\sigma}g_{\sigma\nu}=\delta^{\mu}_\nu-\kappa R^{\mu}_{\nu}(\Gamma),
\label{eq:effective1}
\end{equation}
\begin{equation}
q^{\mu\sigma}g_{\sigma\nu}=\tau\left(\delta^{\mu}_\nu-\kappa T^{\mu}_{\nu}\right),
\label{eq:effective2}
\end{equation}
where we have taken $\Lambda=0$ (i.e. $\lambda=1$) for simplicity, $\tau=\sqrt{g/q}$, $R^{\mu}_{\nu}=q^{\mu\sigma}R_{\sigma\nu}$ and $T^{\mu}_{\nu}=T^{\mu\sigma}g_{\sigma\nu}$. The last two equations can be combined to obtain
\begin{equation}
R^{\mu}_{\nu}(\Gamma)=\tau T^{\mu}_{\nu}+\frac{1-\tau}{\kappa}\delta^\mu_\nu, \quad R(\Gamma)=\tau T+\frac{1-\tau}{\kappa}N,
\end{equation}
where $R=R^\mu_\mu$ and $T=T^\mu_\mu$. The Einstein tensor for $q_{\mu\nu}$ then follows immediately,
\begin{equation}
G^{\mu}_{\nu}(\Gamma)=R^{\mu}_{\nu}(\Gamma)-\frac{1}{2}R\delta^\mu_\nu=\mathcal{T}^{\mu}_\nu,
\end{equation}
where
\begin{equation}
\mathcal{T}^\mu_\nu=\tau T^{\mu}_{\nu}+\mathcal{P}\delta^\mu_\nu, \quad \mathcal{P}=\frac{(\tau-1)(N-2)}{2\kappa}-\frac{1}{2}\tau T.
\end{equation}
$\tau$ can be obtained from $T^\mu_\nu$ by taking determinant on both sides of Eq. (\ref{eq:effective2}). we obtain
\begin{equation}
\tau^2=\tau^N\left[\text{det}\left(\delta^\mu_\nu-\kappa T^\mu_\nu\right)\right]\quad \Rightarrow \tau=\left[\text{det}\left(\delta^\mu_\nu-\kappa T^\mu_\nu\right)\right]^{\frac{-1}{N-2}}.
\end{equation}
Note that, in $(3+1)$ dimensions ($N=4$), for a dust $T^\mu_{\nu}$ (the only non-zero component is $T^t_t=-\rho$, $\rho$ being the energy density), the form of the apparent energy-momentum tensor $\mathcal{T}^\mu_{\nu}$ represents perfect fluid with an effective isotropic pressure given by $\mathcal{P}$ ($\neq 0$) \citep{delsate1}. This complicates the collapse problem in $(3+1)$ dimensions. However, this is not the case in $(2+1)$ dimensions. In $(2+1)$ dimensions ($N=3$), for a dust $T^\mu_{\nu}$, $\tau=1/(1+\kappa\rho)$, $\mathcal{P}=0$ and hence the form of $\mathcal{T}^\mu_{\nu}$ also represents dust, thereby simplifying the field equations to a great extent. Therefore, as a first attempt, we consider the $(2+1)$-dimensional case. Here, we would like to point out that the physical meaning of solutions is given by the physical metric $g_{\mu\nu}$, not by the auxiliary metric $q_{\mu\nu}$; $q_{\mu\nu}$ is introduced for convenience. Therefore, for a dust in $(2+1)$ dimensions, though the effective Einstein equation for $q_{\mu\nu}$ and hence the solution for $q_{\mu\nu}$ is similar to the $(2+1)$ dust solution of GR, the solution for $g_{\mu\nu}$, which determines the physical properties of the solution, will be different from that of GR.

\section{Inhomogeneous dust collapse in (2+1) dimensions}
\label{sec:inhomogeneous}
We assume, respectively, the most general physical and auxiliary metrics of the form
\begin{equation}
ds_g^2=g_{\mu\nu}dx^\mu dx^\nu=-e^{2\alpha(r,t)}dt^2+e^{2\beta(r,t)}dr^2+S^2(r,t)d\theta^2,
\end{equation}
\begin{equation}
ds_q^2=q_{\mu\nu}dx^\mu dx^\nu=-e^{2\nu(r,t)}dt^2+e^{2\psi(r,t)}dr^2+R^2(r,t)d\theta^2.
\end{equation}
For the matter part, we consider a dust form of matter,  whose energy-momentum tensor is given by $T^{\mu \nu}=\rho(r,t)u^\mu u^\nu$, where $\rho(r,t)$ and $u^\mu$ are, respectively, the energy density and four velocity of the dust. The conservation equation $\nabla_\mu T^{\mu t}=0$ gives $\alpha(r,t)=\alpha(t)$. Without loss of generality, we take $\alpha(r,t)=0$. We have another conservation equation $\nabla_\mu T^{\mu r}=0$ which has to be satisfied.

Using field equation (\ref{eq:field_equation1}), we obtain
\begin{equation}
e^{2\nu(r,t)}=\lambda^2=(1+\kappa \Lambda)^2,
\end{equation}
\begin{equation}
\frac{R^2}{S^2}=e^{2\psi-2\beta}=\lambda(\lambda+\kappa \rho)=(1+\kappa \Lambda)(1+\kappa \bar{\rho}),
\end{equation}
where $\bar{\rho}=\rho+\Lambda$. The $tr$-component of the field equation (\ref{eq:field_equation2}) (i.e., $R_{tr}=0$) can be integrated to obtain
\begin{equation}
e^{2\psi(r,t)}=\frac{R'^2(r,t)}{1+f(r)},
\end{equation}
where $f(r)$ is an integration constant. Here, the prime indicates differentiation with respect to $r$. The other components of the field equation (\ref{eq:field_equation2}) are given by
\begin{equation}
\frac{\ddot{R}'}{R'}+\frac{\ddot{R}}{R}=\frac{\lambda^2-1}{\kappa},
\label{eq:component1}
\end{equation}
\begin{equation}
\frac{1-e^{2\beta-2\psi}}{\kappa}=\frac{1}{\lambda^2}\left(\frac{\ddot{R}'}{R'}+\frac{\dot{R}\dot{R}'}{RR'}\right)-\frac{f'}{2RR'},
\label{eq:component2}
\end{equation}
\begin{equation}
\frac{1-e^{2\beta-2\psi}}{\kappa}=\frac{1}{\lambda^2}\left(\frac{\ddot{R}}{R}+\frac{\dot{R}\dot{R}'}{RR'}\right)-\frac{f'}{2RR'},
\label{eq:component3}
\end{equation}
where an overdot denotes differentiation with respect to $t$. Comparing (\ref{eq:component2}) and (\ref{eq:component3}), we get $\ddot{R}'/R'=\ddot{R}/R$. Therefore, from (\ref{eq:component1}), we obtain
\begin{equation}
\ddot{R}+\omega^2R=0,
\label{eq:R_equation}
\end{equation}
where $\omega^2=\frac{\lambda^2-1}{2\kappa}=(1+\frac{\kappa \Lambda}{2})\Lambda$. Therefore, we are left with two equations. Defining $R^2/S^2=e^{2\psi-2\beta}=(1+\kappa \Lambda)(1+\kappa \bar{\rho})=U$, we obtain from (\ref{eq:component3})
\begin{equation}
U(r,t)=\frac{1}{1-\frac{\kappa}{\lambda^2}\left(\frac{\ddot{R}}{R}+\frac{\dot{R}\dot{R}'}{RR'}\right)+\frac{\kappa f'}{2RR'}}.
\label{eq:U_equation}
\end{equation}
Therefore, once we solve Eq. (\ref{eq:R_equation}) for $R(r,t)$, we obtain all the unknown functions. The function $f(r)$ must be related to the initial data. Note that, depending on the cosmological constant $\Lambda$, we have three distinct case, namely,  $\omega^2=0$ ($\Lambda=0$), $\omega^2>0$ ($\Lambda>0$), and $\omega^2<0$ ($\Lambda<0$).

Note that, in obtaining the solutions of the field equations above, we have not put any restriction on sign and value of the cosmological constant so far. However, unlike in GR, in this EiBI theory, the analysis of the collapse with non-zero cosmological constant is quite complicated and involved because of the extra parameter $\kappa$. Therefore, from now onwards, we consider the case of collapse with $\Lambda=0$. In this case $\omega^2=0$. Therefore, the solution of (\ref{eq:R_equation}) is given by $R(t,r)=c_1(r)t+c_2(r)$. Using the freedom of scaling, we choose  initially at $t=0$, 
the radius of the collapsing shells scale as,
\begin{equation}
S(0,r)=\frac{R(0,r)}{\sqrt{1+\kappa\rho_i(r)}}=r,
\end{equation}
where $\rho_i(r)$ is the initial density profile. Therefore, we obtain 
\begin{equation}
R(t,r)=c_1(r)t+r\sqrt{1+\kappa\rho_i(r)}.
\label{eq:R-equation1}
\end{equation}
The function $c_1(r)$ must be chosen in such a way that $S(t,0)=0$. Therefore, we must have $c_1(0)=0$ at the center. Using $U(0,r)=1+\kappa\rho_i(r)$ in (\ref{eq:U_equation}), we obtain
\begin{equation}
f(r)=c_1^2(r)-2\int_0^r\rho_i r dr-\kappa\int_0^r\frac{r^2\rho_i\rho_i'}{1+\kappa\rho_i}dr.
\end{equation}
Therefore, the physical metric becomes
\begin{equation}
ds_g^2=-dt^2+\frac{R'^2-\kappa\frac{R'}{R}\left(\rho_i r+\frac{\kappa r^2}{2}\frac{\rho_i\rho_i'}{1+\kappa\rho_i}\right)}{1+f(r)}dr^2+\left[R^2-\kappa\frac{R}{R'}\left(\rho_i r+\frac{\kappa r^2}{2}\frac{\rho_i\rho_i'}{1+\kappa\rho_i}\right) \right]d\theta^2,
\label{eq:final_metric}
\end{equation}
and the energy density is given by
\begin{equation}
\rho(t,r)=\frac{U-1}{\kappa}=\frac{\rho_i r+\frac{\kappa r^2}{2}\frac{\rho_i\rho_i'}{1+\kappa\rho_i}}{RR'-\kappa\left(\rho_i r+\frac{\kappa r^2}{2}\frac{\rho_i\rho_i'}{1+\kappa\rho_i}\right)}.
\end{equation}
Note that, for $\kappa=0$, the above reduces to the inhomogeneous dust collapse obtained in $(2+1)$-dimensional general relativity \citep{LDC4}.

\section{Spacetime singularities and apparent horizons}
\label{sec:horizons}
The gravitational collapse takes place from a regular initial data given at $t=0$, and any shell labeled by the coordinate $r$ collapses to 
a zero area radius at a time $t=t_s(r)$, i.e., where $S(t_s(r),r)=0$, thereby forming a spacetime singularity. It should be noted that $S(t_s(r),r)=0$ will be satisfied when either (see Eq. \ref{eq:final_metric})
\begin{equation}
R(t_s(r),r)=0, \hspace{0.2cm}\Rightarrow \hspace{0.2cm} t_s(r)=\frac{r\sqrt{1+\kappa\rho_i(r)}}{-c_1(r)}
\label{eq:singularity1}
\end{equation}
or
\begin{equation}
R(t_s(r),r)R'(t_s(r),r)=\kappa\left(\rho_i r+\frac{\kappa r^2}{2}\frac{\rho_i\rho_i'}{1+\kappa\rho_i}\right).
\label{eq:singularity2a}
\end{equation}
Using Eq. (\ref{eq:R-equation1}) and putting $t=0$, Eq. (\ref{eq:singularity2a}) can be rewritten as
\begin{equation}
R(t_s(r),r)R'(t_s(r),r)=\frac{\kappa\rho_i r}{\sqrt{1+\kappa\rho}}R'(0,r).
\label{eq:singularity2}
\end{equation}
In order to avoid occurrence of a shell-crossing singularity, we assume that $R'>0$ throughout the collapse. Therefore, Eq. (\ref{eq:singularity2}) will be satisfied when $\kappa>0$. For $\kappa>0$, Eq. (\ref{eq:singularity2}) will be satisfied before Eq. (\ref{eq:singularity1}) is satisfied. Therefore, the singularity formation time $t_s(r)$ for $\kappa<0$ and $\kappa>0$ are given by the solution of (\ref{eq:singularity1}) and (\ref{eq:singularity2}), respectively. For $\kappa=0$, both these singularity times coincide with that in general relativity collapse. 

At the spacetime singularity formed due to the collapse, both the energy density $\rho(t,r)$ and the curvature scalar (Ricci scalar) diverge for $\kappa>0$. However, for $\kappa<0$, though the Ricci scalar diverges, the energy density remains finite ($\rho(t_s(r),r)=\frac{1}{|\kappa|}$) at the singularity formed due the collapse. This somewhat peculiar feature for the class $\kappa<0$ has been observed in different spherically symmetric, static charged black hole and wormhole spacetimes in this modified gravity theory. We note that this is a markedly different behavior as compared to the general relativity 
scenarios, where curvature scalar divergences typically imply the divergence of
mass-energy density.

The nature of singularities formed during the collapse depends on the formation and dynamics of the apparent horizon. An apparent horizon is represented by $g^{\mu\nu}S_{,\mu} S_{,\nu}=0$. If a given shell labelled by $r$ gets trapped ($g^{\mu\nu}S_{,\mu} S_{,\nu}<0$), and this remains so till the time $t_s(r)$, the result is a black hole singularity. However, if it does not get trapped for all $t\leq t_s(r)$, or becomes untrapped ($g^{\mu\nu}S_{,\mu} S_{,\nu}>0$) as $t\to t_s(r)$, then the result is a formation of naked (timelike) singularity. A timelike singularity is locally naked if future directed outgoing null geodesics emerging from it encounter a trapping region in future. Otherwise, it will be globally naked.

To see how the collapse dynamics evolves from a regular initial data given at $t=0$, we have to choose the functions $c_1(r)$ and $\rho_i(r)$. For $\rho_i(r)$, we choose two types of initial density profile given by
\begin{equation}
\rho_i(r)=\left\{
  \begin{array}{l}
    \rho_0+\rho_2 r^2\\
 \rho_0\frac{\sin \left(\frac{\pi r}{r_b}\right)}{\frac{\pi r}{r_b}}
  \end{array}
\right. ; 0\leq r\leq r_b,
\end{equation}
where $\rho_0$ is the initial density at the center $r=0$ and $r_b$ is given by $\rho_i(r_b)=0$. Concerning $c_1(r)$, we must choose it in such a way that $c_1(0)=0$. We choose two different form of $c_1(r)$. Firstly, we choose $c_1(r)$ in such a way that $f(r)=0$. This gives
\begin{equation}
c_1(r)=-\sqrt{2\int_0^r\rho_i r dr+\kappa\int_0^r\frac{r^2\rho_i\rho_i'}{1+\kappa\rho_i}dr}.
\label{eq:initial_c1_1}
\end{equation}
Note that $c_1(0)=0$. Secondly, we choose $c_1(r)=-r$.
\begin{figure}[ht]
\centering
\includegraphics[scale=0.7]{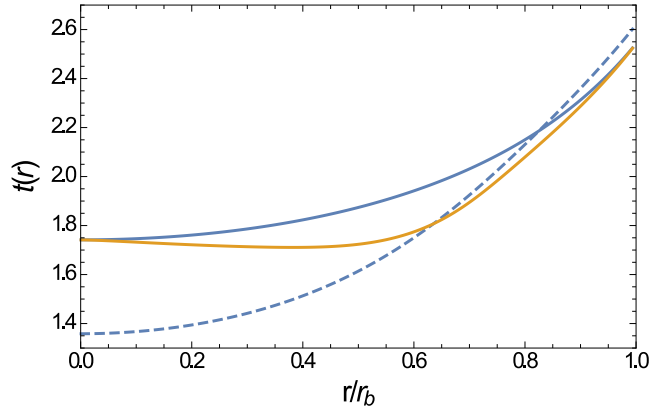}
\caption{Plots showing the singularity time $t_s(r)$ (blue) and the apparent horizon time $t_{ah}(r)$ (orange) for both $\kappa<0$ (solid curves) and $\kappa>0$ (dashed curve). The initial density profile is $\rho_i(r)=\rho_0+\rho_2 r^2$ and $c_1(r)$ given by Eq. (\ref{eq:initial_c1_1}). Here, we have taken $|\kappa|=0.3$, $\rho_0=0.3$, $\rho_2=-0.175$ and $r_b=\sqrt{-\frac{\rho_0}{\rho_2}}$. For $\kappa>0$, no apparent horizon or trapped surfaces form till the singularity time.}
\label{fig:AH1}
\end{figure}
\begin{figure}[ht]
\centering
\includegraphics[scale=0.7]{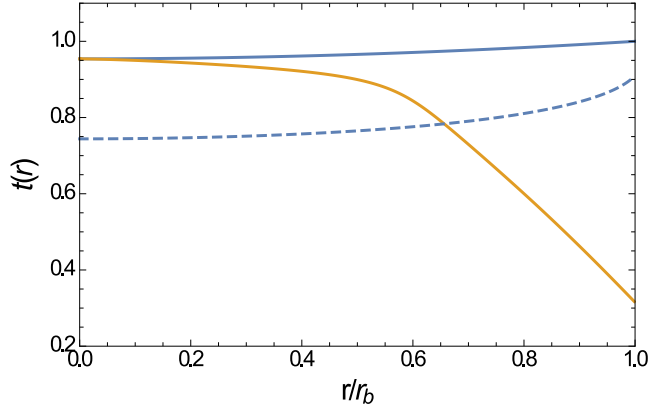}
\caption{Plots showing the singularity time $t_s(r)$ (blue) and the apparent horizon time $t_{ah}(r)$ (orange) for both $\kappa<0$ (solid curves) and $\kappa>0$ (dashed curve). The initial density profile is $\rho_i(r)=\rho_0+\rho_2 r^2$ and $c_1(r)=-r$. Here, we have taken $|\kappa|=0.3$, $\rho_0=0.3$, $\rho_2=-0.175$ and $r_b=\sqrt{-\frac{\rho_0}{\rho_2}}$. For $\kappa>0$, no apparent horizon or trapped surfaces form till the singularity time.}
\label{fig:AH2}
\end{figure}
Figures \ref{fig:AH1} and \ref{fig:AH2} show, for a given initial data, the singularity time $t_s(r)$ and the apparent horizon time $t_{ah}(r)$ for both choices of $c_1(r)$. Note that, for $\kappa<0$, the whole singularity curve is covered since $t_s(r)>t_{ah}(r)$. Therefore, a black hole is always formed as the end state of the collapse for $\kappa<0$. On the other hand, for $\kappa>0$, the whole singularity curve is naked since no apparent horizon or trapped surfaces form till the singularity time. Therefore, for the initial data in Figs. \ref{fig:AH1} and \ref{fig:AH2}, a naked singularity is always formed as the end state of the collapse for $\kappa>0$.

\begin{figure}[ht]
\centering
\subfigure[]{\includegraphics[scale=0.65]{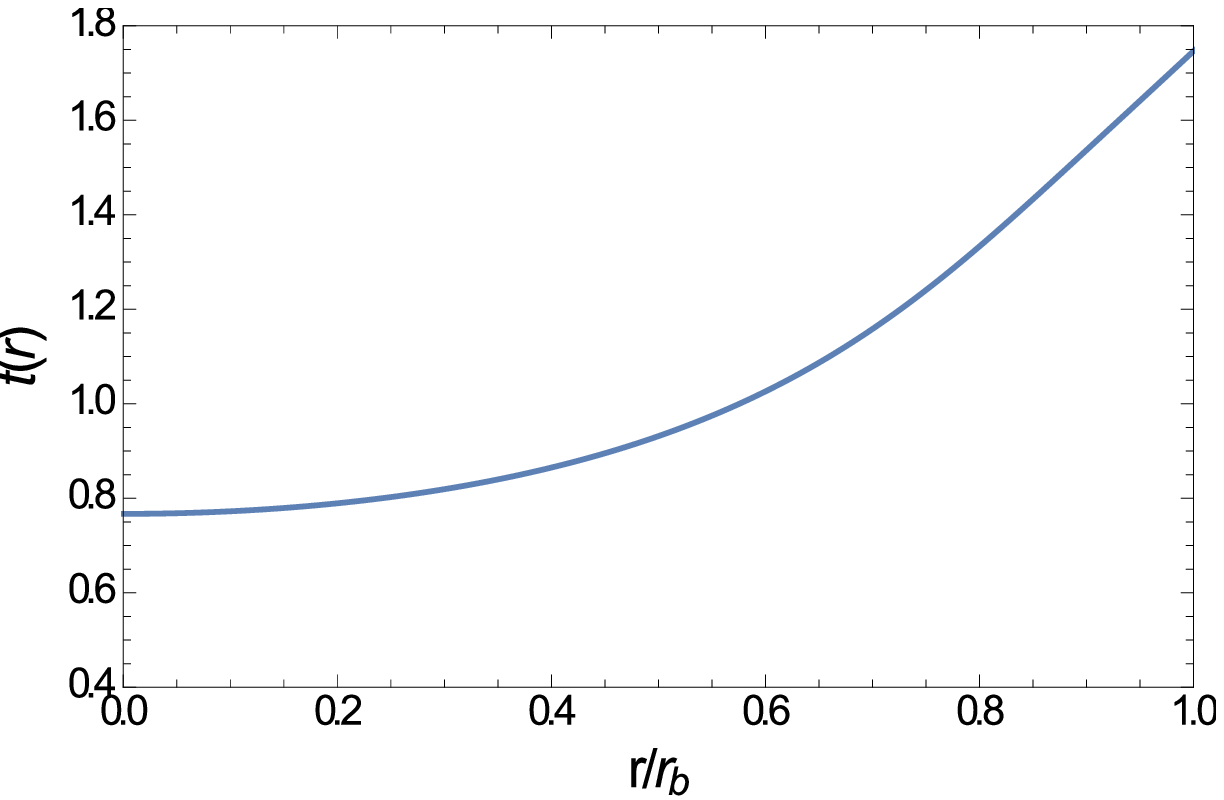}}\hspace{0.1cm}
\subfigure[]{\includegraphics[scale=0.65]{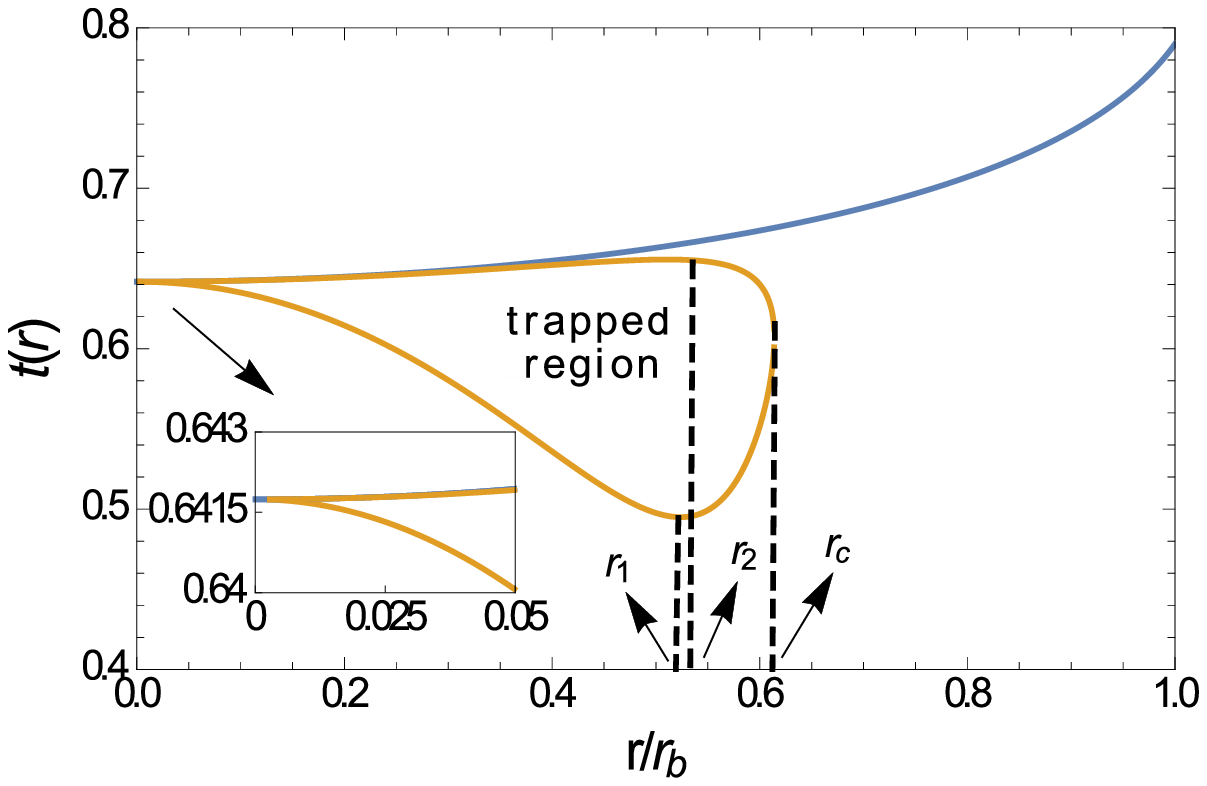}}
\caption{Plots showing the singularity time $t_s(r)$ (blue) and the apparent horizon time $t_{ah}(r)$ (orange) for (a) $c_1(r)$ given by Eq. (\ref{eq:initial_c1_1}) and (b) $c_1(r)=-r$. The initial density profile is $\rho_i(r)=\rho_0+\rho_2 r^2$. Here, we have taken $\kappa=0.3$, $\rho_0=0.7$, $\rho_2=-0.175$ and $r_b=\sqrt{-\frac{\rho_0}{\rho_2}}$.}
\label{fig:AH_positive}
\end{figure}

However, in the $\kappa>0$ case, with a given different initial data (e.g. with a higher initial central density), apparent horizons may form for the second choice of $c_1(r)$, i.e., for $c_1(r)=-r$. Figure \ref{fig:AH_positive} shows such an example where the initial central density is taken to be higher than that used in Figs. \ref{fig:AH1} and \ref{fig:AH2}.
Note that, with the higher initial central density, still an apparent horizon does not form for the choice of $c_1(r)$ given by $f(r)=0$. However, with the same higher initial central density, apparent horizons do form for the choice $c_1(r)=-r$, and the dynamics of the apparent horizons in this case are different from that obtained in the $\kappa<0$ case. In this case, an apparent horizon is first formed at $r=r_1$. As the collapse evolves, this apparent horizon then splits in two-- one traveling inward and other traveling outward. A trapped region is formed between the two apparent horizons. As the collapse evolves further, the outer apparent horizon takes a turn at $r=r_c$ and starts traveling inward. Whereas, the inner apparent horizon approaches the $r=0$ shell and collapses to zero radius. However, at the same time, another apparent horizon is formed at $r=0$. This apparent horizon travels outward and annihilate with the outer one at $r=r_2$. The whole singularity curve lies outside the trapping region between the two apparent horizon. A shell labeled by $r<r_c$ gets trapped at $t=t_{ah}(r)$ given by the lower orange curve and then gets untrapped (at $t=t_{ah}(r)$ given by the upper orange curve) before it collapses to zero radius. Therefore, the singularities formed due to collapse of these shells are timelike and hence, are naked. The apparent horizon does not form for $r>r_c$. These shells outside $r=r_c$ do not get trapped for all $t\leq t_s(r)$. Hence, the singularities formed out of the collapse of these shells are timelike and naked. For the second choice of the initial density profile $\rho_i(r)$, similar conclusions hold.

From the above analysis for the two sets of initial conditions, it is clear that the singularity formed out of the collapse is covered for $\kappa<0$ and naked for $\kappa>0$. This can also be shown in a general way by evaluating $g^{\mu\nu}S_{,\mu} S_{,\nu}$ along the singularity curve. The expression for $g^{\mu\nu}S_{,\mu} S_{,\nu}$ is complicated. However, it can be shown that $\lim_{t\to t_s(r)} g^{\mu\nu}S_{,\mu} S_{,\nu}\to-\infty$ for $\kappa<0$. Therefore, the singularity for $\kappa<0$ is always trapped, spacelike and hence covered. On the other hand $\lim_{t\to t_s(r)} g^{\mu\nu}S_{,\mu} S_{,\nu}\to \infty$ for $\kappa>0$. Therefore, the singularity for $\kappa>0$ is always timelike and hence naked.

\section{Homogeneous dust collapse}
\label{sec:homogeneous}
Let us now consider the homogeneous case. We set $\rho_i(r)=\rho_0$. Therefore, for the choice $f(r)=0$, i.e., for $c_1(r)=-\sqrt{\rho_0}r$, the metric and the energy density become, respectively,
\begin{equation}
ds_g^2=-dt^2+a^2(t)\left[dr^2+r^2d\Omega^2\right]
\end{equation}
and $\rho(t)=\frac{\rho_0}{a^2}$, where
\begin{equation}
a(t)=\sqrt{(-\sqrt{\rho_0}t+\sqrt{1+\kappa\rho_0})^2-\kappa\rho_0}.
\end{equation}
Note that, for $\kappa>0$, all the shells collapse simultaneously at time given by
\begin{equation}
t=t_s=\frac{(\sqrt{1+\kappa\rho_0}-\sqrt{\kappa\rho_0})}{\sqrt{\rho_0}}.
\end{equation}
In this case, the apparent horizon time is given by
\begin{equation}
t_{ah}(r)=\frac{1}{\sqrt{\rho_0}}\left[\sqrt{1+\kappa\rho_0}-\frac{\sqrt{\kappa\rho_0}}{\sqrt{1-\rho_0 r^2}}\right]\leq t_s.
\end{equation}
Therefore, the singularity formed in the homogeneous collapse for $\kappa>0$ is always covered for $f(r)=0$. Note that, in the inhomogeneous case, the singularity for $\kappa>0$ was always naked. This implies that the inhomogeneity plays an important role in making the singularity visible for $\kappa>0$.

On the other hand, for $\kappa<0$, all the collapsing shells undergo a bounce at $t=t_b=\sqrt{\frac{1+\kappa\rho_0}{\rho_0}}$, thereby avoiding the formation of singularity. In the inhomogeneous case, for $\kappa<0$, we always had formation of black hole singularity out of the collapse. However, the homogeneous case for $\kappa<0$ gives bouncing solution. Therefore, it can be concluded that, under small inhomogeneous perturbation, singularity avoidance through the homogeneous bounce will be unstable, leading to the formation of spacetime singularity. Similar bouncing solutions have been observed in homogeneous cosmology in this gravity theory. Therefore, our result suggests that these bouncing cosmological solutions may be unstable under small inhomogeneous perturbation. In fact, it has been found that, for $\kappa<0$, linear perturbations of the homogeneous bouncing cosmological solutions in four dimensional EiBI gravity are unstable \citep{perturbation1,perturbation2}. For the other choice of $c_1(r)$, i.e., for $c_1(r)=-r$, similar conclusion holds.

To illustrate further the qualitative difference between the inhomogeneous and the homogeneous collapse for $\kappa<0$, let us consider the physical area radius $S(t,r)$ given by
\begin{equation}
S^2(t,r)=\frac{R}{R'}\left[RR'-\kappa\left(\rho_i r+\frac{\kappa r^2}{2}\frac{\rho_i\rho_i'}{1+\kappa\rho_i}\right) \right],
\label{eq:S-equation}
\end{equation}
where $R(t,r)$ is given by (\ref{eq:R-equation1}). Let us now consider the $f(r)=0$ and $\rho_i=\rho_0+\rho_2r^2$ case. We further consider that the inhomogeneity is very small, i.e., $\big|\frac{\rho_2}{\rho_0}\big|r^2\ll 1$. For $\kappa<0$, the term inside the square bracket in the above equation is always positive. Therefore, the physical area radius $S(t,r)$ goes to zero only when the auxiliary area radius $R(t,r)$ goes to zero and $R'\neq 0$. For small inhomogeneity, i.e., for $\big|\frac{\rho_2}{\rho_0}\big|r^2\ll 1$, we have
\begin{equation}
\frac{R}{R'}\simeq r\frac{-\sqrt{\rho_0}t\left[1+\frac{\rho_2}{4\rho_0}\left(\frac{1+2\kappa\rho_0}{1+\kappa\rho_0}\right)r^2\right]+\sqrt{1+\kappa\rho_0}}{-\sqrt{\rho_0}t\left[1+\frac{3\rho_2}{4\rho_0}\left(\frac{1+2\kappa\rho_0}{1+\kappa\rho_0}\right)r^2\right]+\sqrt{1+\kappa\rho_0}+\frac{\kappa\rho_2 r^2}{\sqrt{1+\kappa\rho_0}}}.
\end{equation}
Now, at $t=t_{s/b}(r)$ given by $R(t_{s/b}(r),r)=0$, we have
\begin{equation}
R'(t_{s/b}(r),r)\simeq \frac{1}{\sqrt{1+\kappa\rho_0}}\left(-\frac{\rho_2}{\rho_0}\right)r^2,
\end{equation}
which is positive and non-zero for $\rho_2\neq 0$. Therefore, in the inhomogeneous case $(\rho_2\neq 0)$ with $\kappa<0$, $R(t_{s/b}(r),r)=0$ implies $S(t_{s/b}(r),r)=0$, thereby implying a spacetime singularity with $t_{s/b}(r)=t_s(r)$ being the singularity time. However, in the homogeneous case $(\rho_2=0)$, both $R$ and $R'$ become zero at $t=t_{s/b}(r)$, and $\frac{R}{R'}=r$ for all $t$ including $t=t_{s/b}(r)$. Therefore, in the homogeneous case with $\kappa<0$, $R(t_{s/b}(r),r)=0$ does not imply $S(t_{s/b}(r),r)=0$. In fact, it is clear from Eq. (\ref{eq:S-equation}) that, in this case, the term within the square bracket becomes minimum at $t=t_{s/b}(r)$ ($=\sqrt{\frac{1+\kappa\rho_0}{\rho_0}}$), thereby giving a bounce with $t_{s/b}=t_{b}=\sqrt{\frac{1+\kappa\rho_0}{\rho_0}}$ being the time when the bounce occurs. This explains why a small inhomogeneity makes the homogeneous bounce unstable, thereby making it singular.

\section{Conclusions}
\label{sec:conclusion}
In this paper, from the perspective of cosmic censorship, we have investigated the collapse of a dust cloud in (2+1) dimension in Eddington-inspired Born-Infeld gravity (EiBI). We have studied the formation of black holes (BH) and naked singularities (NS) as the final states of collapse and compared the results with those of general relativity (GR). It is found that, as opposed to the general relativistic situation where the outcome of dust collapse in $(2+1)$ dimensions is always a naked singularity \citep{LDC5}, the formation of naked singularity can be avoided in EiBI gravity for negative Born-Infeld theory parameter $\kappa$. The inhomogeneous dust collapse with negative $\kappa$ always leads to the formation of a black hole singularity. On the other hand, the inhomogeneous dust collapse with positive $\kappa$ always leads to the formation of a naked singularity. This indicates that a $(3+1)$ dimensional generalization of these results could be useful and worth examining. Finally, using the results here, we have shown that the singularity avoidance through homogeneous bounce in cosmology in this modified gravity is not stable.

Here, as pointed out earlier, we have analysed the collapse with zero cosmological constant. Unlike in GR, in EiBI theory, the analysis of the collapse with non-zero cosmological constant is complicated and involved because of the extra parameter $\kappa$. It will be interesting to see whether and to what extent the collapse with non-zero cosmological constant qualitatively differs from that with zero cosmological constant. This case has to be studied separately in some detail. We hope to address this in future.

\end{document}